\documentclass[twocolumn,pre,superscriptaddress,amsmath,amssymb]{revtex4}

\usepackage{graphicx}
\usepackage{dcolumn}
\usepackage{bm}
\usepackage{float}
\usepackage{color}
\usepackage{soul}
\usepackage{natbib}

\begin{document}


\title{Structural Properties of Hyperuniform Networks}
\author{Eli Newby}
\affiliation{Department of Physics, Pennsylvania State University University Park, PA 16802}

\author{Wenlong Shi}
\affiliation{Materials Science and Engineering, Arizona State
University, Tempe, AZ 85287}

\author{Yang Jiao}
\email[correspondence sent to: ]{yang.jiao.2@asu.edu}
\affiliation{Materials Science and Engineering, Arizona State
University, Tempe, AZ 85287} \affiliation{Department of Physics,
Arizona State University, Tempe, AZ 85287}

\author{Reka Albert}
\email[correspondence sent to: ]{rza1@psu.edu}
\affiliation{Department of Physics, Pennsylvania State University University Park, PA 16802}

\author{Salvatore Torquato}
\email[correspondence sent to: ]{torquato@electron.princeton.edu}
\affiliation{
Department of Chemistry, Princeton University, Princeton, New Jersey 08544, USA}
\affiliation{
Department of Physics,  Princeton University, Princeton, New Jersey 08544, USA}
\affiliation{
Princeton Institute  of Materials, Princeton University, Princeton, New Jersey 08544, USA}
\affiliation{
Program in Applied and Computational Mathematics, Princeton University, Princeton, New Jersey 08544, USA}



\date{\today}

\begin{abstract}
Disordered hyperuniform many-particle systems are recently discovered exotic states of matter, characterized by a complete suppression of normalized infinite-wavelength density fluctuations, as in perfect crystals, and lack of conventional long-range order, as in liquids and glasses. In this work, we begin a program to quantify the structural properties of nonhyperuniform and hyperuniform networks. In particular, large two-dimensional (2D) Voronoi networks (graphs) containing approximately 10,000 nodes are created from a variety of different point configurations, including the antihyperuniform hyperplane intersection process (HIP), nonhyperuniform Poisson process, nonhyperuniform random sequential addition (RSA) saturated packing, and both non-stealthy and stealthy hyperuniform point processes. We carry out an extensive study of the Voronoi-cell area distribution of each of the networks through determining multiple metrics that characterize the distribution, including their higher-cumulants (i.e., skewness $\gamma_1$ and excess kurtosis $\gamma_2$). We show that the HIP distribution is far from Gaussian, as evidenced by a high skewness ($\gamma_1 = 3.16$) and large positive excess kurtosis ($\gamma_2 = 16.2$). The Poisson (with $\gamma_1 = 1.07$ and $\gamma_2 = 1.79$) and non-stealthy hyperuniform (with $\gamma_1 = 0.257$ and $\gamma_2 = 0.0217$) distributions are Gaussian-like distributions, since they exhibit a small but positive skewness and excess kurtosis. The RSA (with $\gamma_1 = 0.450$ and $\gamma_2 = -0.0384$) and the highest stealthy hyperuniform distributions (with $\gamma_1 = 0.0272$ and $\gamma_2 = -0.0626$) are also non-Gaussian because of their low skewness and negative excess kurtosis, which is diametrically opposite non-Gaussian behavior of the HIP. The fact that the cell-area distributions of large, finite-sized RSA and stealthy hyperuniform networks (e.g., with $N \approx 10,000$ nodes) are narrower, have larger peaks, and smaller tails than a Gaussian distribution implies that in the thermodynamic limit the distributions should exhibit compact support, consistent with previous theoretical considerations. Moreover, we compute the Voronoi-area correlation functions $C_{00}(r)$ for the networks, which describe the correlations between the area of two Voronoi cells separated by a given distance $r$ [M. A. Klatt and S. Torquato, Phys. Rev. E {\bf 90}, 052120 (2014)]. We show that the correlation functions $C_{00}(r)$ qualitatively distinguish the antihyperuniform, nonhyperuniform and hyperuniform Voronoi networks considered here. Specifically, the antihyperuniform HIP networks possess a slowly decaying $C_{00}(r)$ with large positive values, indicating large fluctuations of Voronoi cell areas across scales. While the nonhyperuniform Poisson and RSA network possess positive and fast decaying $C_{00}(r)$, we find strong anticorrelations in $C_{00}(r)$ (i.e., negative values) for the hyperuniform networks. The latter indicates that the large-scale area fluctuations are suppressed by accompanying large Voronoi cells with small cells (and vice versa) in the systems in order to achieve hyperuniformity. In summary, we have shown that cell-area distributions and pair correlations functions of Voronoi networks enable one to distinguish quantitatively antihyperuniform, standard nonhyperuniform and hyperuniform networks from one another.

\end{abstract}

\maketitle


\section{Introduction}



Disordered hyperuniform many-particle systems in $d$-dimensional Euclidean space $\mathbb{R}^d$ are exotic amorphous states in which the normalized density fluctuations are completely suppressed at infinite wavelengths \cite{To03, To18a}. In the last two decades, such disordered systems have been shown to arise in a variety of contexts in the physical, mathematical and biological contexts \cite{To03,To18a,Do05,Za09,Ji14,To16a,Leseur16, hexner2017noise, hexner2017enhanced, weijs2017mixing, torquato2019hidden, chen2021multihyperuniform, Zh20, salvalaglio2020hyperuniform, Ch21, sakai2022quantum, puig2022anisotropic, ge2023hidden, Philcox2023, zhang2023approach, BRO_origin, liu2024universal, sanchez2023disordered}. Disordered hyperuniform materials are attracting great attention because they possess many unique optical, mechanical and transport properties \cite{Fl09,Ma13,FL13,We15, xu2017microstructure, wu2017effective, Ru19,Lei2019, zhang2019experimental, cheron2022wave, tavakoli2022over, granchi2022near, tang2022soft, chen2022disordered, chen2023disordered, zhuang2024vibrational}.

A \textit{hyperuniform} system, disordered or not, can be defined in two equivalent ways. A hyperuniform many-particle systems is one in which the number variance of particles within a spherical observation window of radius $R$ grows slower than the observation window volume $R^d$ for large $R$ \cite{To03,To18a}, i.e.,
\begin{equation}
\lim_{R\rightarrow\infty}\sigma_N^2(R)/v_1(R) = 0
\end{equation}
where $\sigma_N^2(R)\equiv\langle N(R)^2\rangle - \langle N(R)\rangle^2$ is the variance in number of points (brackets indicate the ensemble average) and $v_1(R) \sim R^d$ is the volume of a $d$-dimensional sphere or radius $R$. Equivalently, a many-particle system is hyperuniform if its structure factor $S(\mathbf{k})$ tends to zero as the wavenumber $k\equiv|\mathbf{k}|$ tends to zero, i.e.,
\begin{equation}
    \lim_{|\mathbf{k}|\rightarrow0}S(\mathbf{k})=0.
\end{equation}
The small-$k$ scaling behavior of $S(k) \sim k^\alpha$ determines the large-$R$
asymptotic behavior of $\sigma_N^2(R)$, based on which all hyperuniform
systems, disordered or not, can be categorized into three classes:
$\sigma_N^2(R) \sim R^{d-1}$ for $\alpha>1$ (class I); $\sigma_N^2(R)
\sim R^{d-1}\ln(R)$ for $\alpha=1$ (class II); and $\sigma_N^2(R)
\sim R^{d-\alpha}$ for $0<\alpha<1$ (class III) \cite{To18a}. Consequently, hyperuniform systems encompass all crystals and quasicrystals \cite{To03, To18a}. A special subset of class-I hyperuniform systems possess a zero structure factor for a range of wavenumbers around the origin, i.e., $S(k) = 0$ for $k<K^*$ (excluding the forward scattering), which are referred as stealthy hyperuniform systems. Stealthy hyperuniform systems include all crystals and certain special disordered systems, which are characterized by a parameter $\chi \in (0, 0.5)$ reflecting the fraction of the constrained degrees of freedom in the system \cite{To15, Zh15a, Zh15b, Zh17}, see Sec. II for more details. Increasing $\chi$ also leads to an increasing of local order in the system.



Although this exotic state of matter continues to be identified in more fields and their unique properties are being further illuminated, the preponderance of previous work has focused on hyperuniform particle configurations, two-phase media and random fields \cite{To03, ref2, To16a, To18a}. On the other hand, only a few studies on hyperuniform networks \cite{klatt2014characterization, salvalaglio2024persistent} have been carried out. Here, we focus on the ``spatial networks'' derived from point configurations, i.e., graphs whose nodes are embedded in a space with a metric (e.g., a Euclidean space) \cite{Barthelemy2011,Daqing2011}. In particular, we construct networks using the Voronoi tessellations of point patterns \cite{Aurenhammer1991,To02a,Lazar2022}.

A Voronoi tessellation in 2D is a partition of the plane by convex polygonal cells which are referred to as Voronoi cells \cite{To02a}. The Voronoi cell associated with a point is defined to be the region of space nearer to this point at than to any other point in the point pattern (see Sec. II.B for details). We convert the tessellation into a network by specifying the vertices of the Voronoi polygons as nodes and the edges of the polygons as edges of the network. We note there are multiple ways of converting a point pattern into a spatial network \cite{Yook2002,Bullmore2009,Farahani2013,Papadopolous2018,Duchemin2023,Penrose2003}, and each was found to have both pros and cons. We choose Voronoi networks because such networks are widely applied in numerous different fields, including biology, meteorology, metallurgy, crystallography, forestry, ecology, archaeology, geology, geography, astrophysics, physics, computer science, and engineering \cite{To02a}. Moreover, previous theoretical studies have established analytical results for such systems that can be used as benchmarks for our current study \cite{Zh15a,Zhang2013}.





We comprehensively analyze the networks constructed from the Voronoi tessellation of the point patterns. First, we perform an analysis of the area distribution of Voronoi cells to recapitulate previously found results regarding the size of holes in the point pattern \cite{Zh16,Zh17,Ghosh2018,torquato2021local,Wang2023}. In particular, we calculate the higher-cumulants, including skewness $\gamma_1$ and excess kurtosis $\gamma_2$ associated with the Voronoi tessellations of large, finite-sized two-dimensional systems containing $N\approx 10,000$ points built from hyperplane intersection process (HIP) \mbox{\cite{Chiu2013}}, Poisson, random sequential addition (RSA) \mbox{\cite{Torquato2006,Zhang2013}}, and both non-stealthy \mbox{\cite{wang2023equilibrium, wang2024designer}} and stealthy \mbox{\cite{Uc04,Ba08,To15}} point processes. The higher order cumulants of each distribution reveal the shape of the cell-area distribution and compare the probability density function to a Gaussian distribution (i.e., $\gamma_1 = 0$, $\gamma_2 = 0$).


Our analysis reveals that the Poisson and non-stealthy hyperuniform cell-area distributions are Gaussian-like because they possess small, positive skewness and excess kurtosis values (i.e., $\gamma_1 = 1.07,~ 0.257$ and $\gamma_2 = 1.79, ~0.0217$ for the Poisson and non-stealthy hyperuniform distributions respectively). In contrast, the antihyperuniform HIP system has a large positive skewness ($\gamma_1 = 3.16$) and excess kurtosis ($\gamma_2 = 16.2$), indicating it is far from Gaussian. The RSA and high-$\chi$ stealthy hyperuniform systems are also non-Gaussian because they have low skewness and negative excess kurtosis (i.e., $\gamma_1 = 0.450, ~0.0492, ~0.0272$ and $\gamma_2 = -0.0383, ~ -0.119, ~ -0.0626$ for the RSA, stealthy $\chi = 0.40$, and stealthy $\chi = 0.49$ respectively), yet they possess diametrically opposite non-Gaussian behavior of the HIP systems. Moreover, the low-$\chi$ stealthy hyperuniform systems are not Gaussian-like because they are also very peaked and their right tail decreases faster than a Gaussian curve. However, this is not directly readable from their skewness and excess kurtosis values due to a left tail in their distributions.





We note that the Voronoi networks analyzed here are necessarily finite in size, even if relatively large with $N \approx 10,000$ nodes. It is clear that distributions associated with finite systems will always appear to possess finite support, even if the support becomes infinite (such as a Poisson distribution) in the thermodynamic limit (i.e., $N \rightarrow \infty$) \cite{Zh15a}. This is because the probability of observing arbitrarily large holes of the order of the system size is vanishing small. Therefore, in order to correctly ascertain the behaviors of infinite networks, the tail behavior of the simulated finite-sized distributions are analyzed. Systems possessing infinite support in the thermodynamic limit (e.g., Poisson) can be inferred from an exponential or power-law decaying tail in the distributions associated with sufficiently large, yet finite-sized systems. Similarly, one can infer compact support for RSA and all stealthy cases from our finite-sized simulations in the same way consistent with the known
theoretical results for these models \cite{Zhang2013,Zh15a}.

To further characterize the Voronoi networks, we compute the Voronoi-area correlation functions $C_{00}(r)$ for the networks \cite{klatt2014characterization}. In particular, $C_{00}(r)$ is defined as the correlation between the volume of two Voronoi cells given that the corresponding centers separated by distance $r$ (see Sec. II.B for details). We observe strong anticorrelations in $C_{00}(r)$ (i.e., negative values) for the hyperuniform networks, consistent with previous analysis of the hyperuniform MRJ sphere packings in 3D \cite{klatt2014characterization}. On the other hand, the antihyperuniform HIP networks possess a slowly decaying $C_{00}(r)$ with large positive values, indicating large fluctuations of Voronoi cell areas across scales, while the nonhyperuniform Poisson and RSA network possess positive and much faster decaying $C_{00}(r)$. These results indicate that the large-scale area fluctuations in the hyperuniform networks are effectively suppressed by accompanying large Voronoi cells with small cells (and vice versa), in contrast to the nonhyperuniform networks in which the Voronoi cells with different areas do not possess any notable correlations. To summarize, we find the correlation functions $C_{00}(r)$ qualitatively distinguish the antihyperuniform, nonhyperuniform and hyperuniform Voronoi networks considered here.

The rest of the paper is organized as follows: In Sec. II, we provide definitions of hyperuniform as well as the Voronoi-area correlation functions. In Sec. III, we present our results, including the comprehensive Voronoi distribution analysis as well as the correlation function analysis of all seven network systems. In Sec. IV, we provide concluding remarks and discuss future directions. 


\section{Definitions}
\subsection{Structure Factor and Hyperuniformity}

Point configurations in $d$-dimensional Euclidean space $\mathbb{R}^d$ are fully spatially characterized by an infinite set of $n$-particle correlation functions $\rho_n(\mathbf{r}_1,\dots,\mathbf{r}_n)$, which are proportional to the probability of finding $n$ particles at the positions $\mathbf{r}_1,\dots,\mathbf{r}_n$ \cite{torquato2002random}.
For statistically homogeneous systems, $\rho_1(\mathbf{r}_1)=\rho$, and $\rho_2(\mathbf{r}_1,\mathbf{r}_2)=\rho^2 g_2(\mathbf{r})$, where $\mathbf{r}=\mathbf{r}_1-\mathbf{r}_2$, and $g_2(\mathbf{r})$ is the pair correlation function.
If the system is also statistically isotropic, then $g_2(\mathbf{r})=g_2(r)$, where $r=|r|$.
The ensemble-averaged structure factor $S(\mathbf{k})$ is defined as
\begin{equation}
    S(\mathbf{k})=1+\rho\Tilde{h}(\mathbf{k})
\end{equation}
where $\Tilde{h}(\mathbf{k})$ is the Fourier transform of the total correlation function $h(\mathbf{r})=g_2(\mathbf{r})-1$.

For a single periodic point configuration with $N$ particles at positions $\mathbf{r}^N = (\mathbf{r}_1,\dots,\mathbf{r}_N)$ within a fundamental cell $F$ of a lattice $\Lambda$, the scattering intensity $\mathbb{S}(\mathbf{k})$ is given by

\begin{equation}\label{eq:Skcomp}
    \mathbb{S}(\mathbf{k}) = \frac{|\sum_{j=1}^N \textrm{exp}(-i\mathbf{k}\cdot\mathbf{r}_j)|^2}{N}.
\end{equation}
In the thermodynamic limit, an ensemble of $N$-particle configurations in $F$ is related to $S(\mathbf{k})$ by
\begin{equation}
    \lim_{N,V_F\rightarrow\infty}\langle \mathbb{S}(\mathbf{k})\rangle = (2\pi)^d \rho \delta(\mathbf{k}) + S(\mathbf{k}),
\end{equation}
where $V_F$ is the volume of the fundamental cell and $\delta$ is the Dirac delta function \cite{To03}.
For finite-$N$ simulations under periodic boundary conditions, Eq. \ref{eq:Skcomp} is used to compute $S(\mathbf{k})$ directly by averaging over configurations.


Consider systems characterized by a structure factor with a radial power law in the vicinity of the origin,
\begin{equation}
    S(\mathbf{k})\sim|\mathbf{k}|^{\alpha}\;\textrm{for}\;|\mathbf{k}|\rightarrow0.
\end{equation}
For {\it hyperuniform} systems, $\alpha > 0$, which is referred to as the hyperuniformity exponent. A (standard) {\it nonhyperuniform} system is one for which $\alpha = 0$, i.e., S({\bf k}) approaches a non-zero constant in the zero-wavenumber limit. An {\it antihyperuniform} system is one possessing a diverging S({\bf k}) in the zero-wavenumber limit, i.e., with $\alpha < 0$.

For hyperuniform systems, the specific value of $\alpha$ determines large-$R$ scaling behaviors of the number variance \cite{To03, To18a}, according which all hyperuniform systems can be categorized into three different classes:
\begin{equation}\label{eq:classes}
    \sigma^2_N(R)\sim
    \begin{cases}
    R^{d-1}&\alpha > 1, \textrm{class I}\\
    R^{d-1}\textrm{ln}(R)&\alpha = 1, \textrm{class II}\\
    R^{d-\alpha}&\alpha < 1, \textrm{class III}.
    \end{cases}
\end{equation}
Classes I and III are the strongest and weakest forms of hyperuniformity, respectively. Class I media include all crystal structures \cite{To03}, many quasicrystal structures \cite{Og17} and exotic disordered systems \cite{Za09, Ch18a}. Examples of Class II systems include some quasicrystal structures \cite{Og17}, perfect glasses \cite{zhang2017classical}, and maximally random jammed packings \cite{Do05, Za11a, Ji11, Za11c, Za11d}. Examples of Class III systems include classical disordered ground states \cite{Za11b}, random organization models \cite{He15}, perfect glasses \cite{zhang2017classical}, and perturbed lattices \cite{Ki18}; see Ref. \cite{To18a} for a more comprehensive list of systems that fall into the three hyperuniformity classes.


Stealthy hyperuniform systems are a special subset of class-I hyperuniform systems possessing a zero structure factor for a range of wavevectors around the origin, i.e.,
\begin{equation}
S({\bf k}) = 0, \quad\text{for}\quad {\bf k} \in \Omega,
\label{eq_stealthy}
\end{equation}
excluding the forward scattering. Stealthy hyperuniform systems include all crystals and certain special disordered systems, which are characterized by a parameter $\chi$ reflecting the fraction of the constrained degrees of freedom in the system \cite{To15, Zh15a, Zh15b, Zh17}, i.e.,
\begin{equation}
    \chi = N_\Omega/(N-d)
    \label{eq_chi_ratio}
\end{equation}
which is the fraction of constrained degree of freedom in the systems. We note $d$ degrees of freedom associated with the trivial overall translation of the entire system are subtracted in Eq. (\ref{eq_chi_ratio}). In the case of point configurations, it has been shown that increasing $\chi$ leads to increased degree of order in the systems \cite{To18a, Ba08}. Disordered stealthy hyperuniform point configurations can be generalized using the so-called ``collective coordinates'' approach \cite{Ba08, Zh15a, Zh15b}.

\subsection{Voronoi Network and Voronoi-Area Correlation Function}



\begin{figure*}[htpb]
    \includegraphics[width=\textwidth]{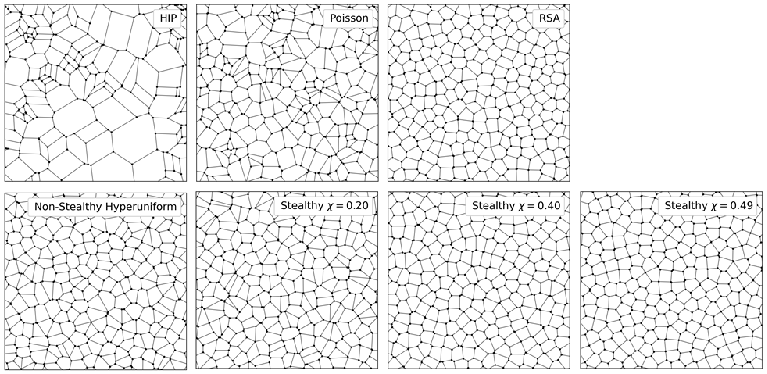}
\caption{Snapshot of the Voronoi tessellations of each of the different point processes. The systems in the top row are not hyperuniform, and the systems in the bottom row are hyperuniform.}
\label{fig_1}
\end{figure*}


\begin{figure*}[htpb]
    \includegraphics[width=\textwidth]{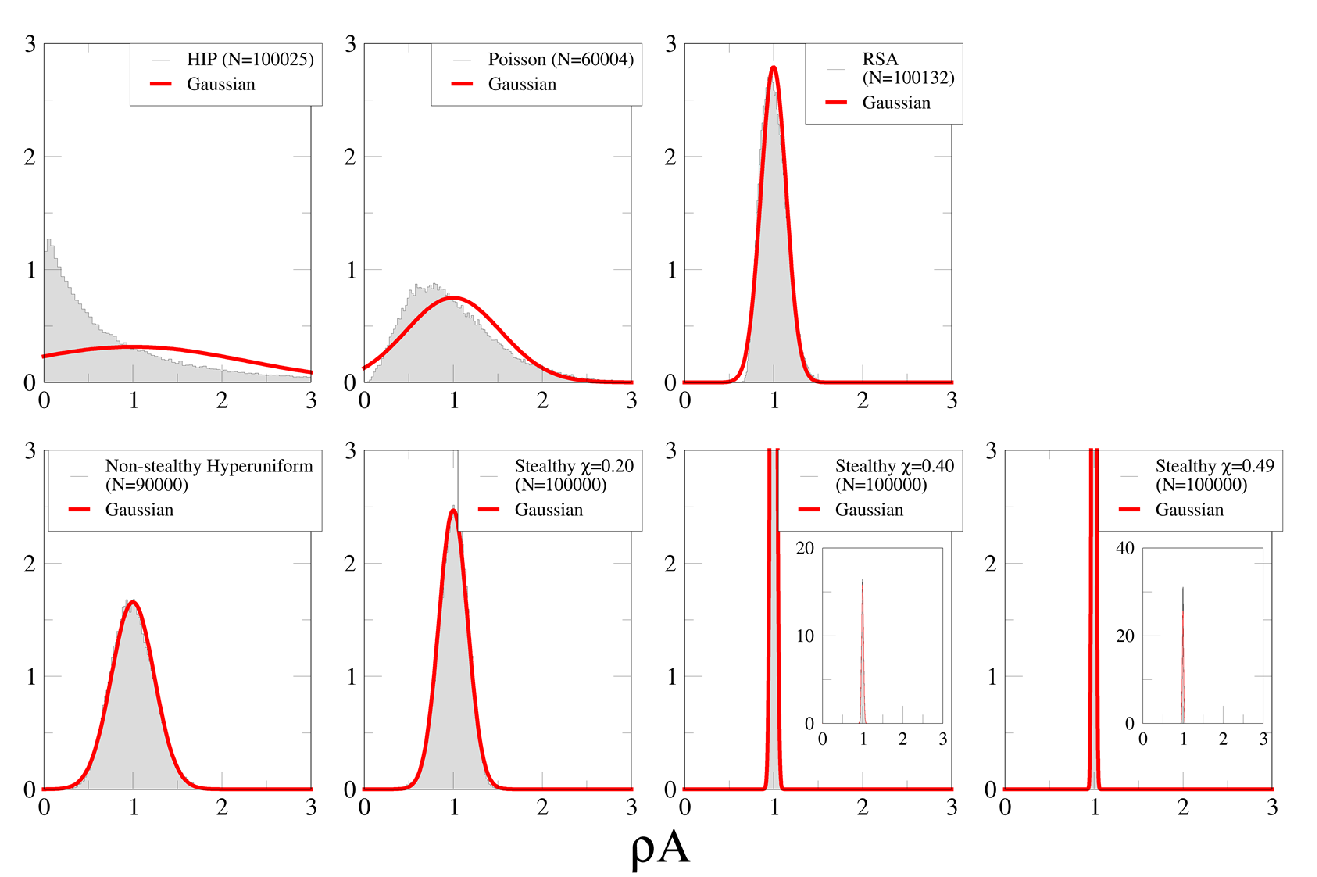}
\caption{Graphs of the normalized Voronoi cell-area probability density functions (PDF) calculated from Voronoi tessellations of approximately 10,000 faces associated with various systems. We normalize the cell areas by the density of the system $\rho$ to make the systems comparable to each other. A Gaussian PDF with the same mean and standard deviation is plotted on top of each distribution (red line). To allow for all the distributions to be discernible, we cutoff both the x and y axes at 3. For the stealthy $\chi = 0.40$ and $\chi = 0.49$ we include an insert that shows the entire distribution. }
\label{fig_2}
\end{figure*}

A network (or graph) $\mathcal{G}(\mathcal{N},\mathcal{E})$ is defined by a set of nodes $\mathcal{N}$ and its set of connecting $M$ edges $\mathcal{E}$. An edge indicates a local interaction or relationship between a pair of nodes. The network is purely defined by the set of nodes and edges, so it can represent an abstraction of the physical system. In this work, we mainly focus on 2D Voronoi networks associated with a configuration of $n$ points ${\bf r}_1, {\bf r}_2, \ldots, {\bf r}_n$ \cite{To02a}. Associated with the $i$-th point at ${\bf r}_i$ is the Voronoi cell, which is defined to be the region of space nearer to the point at ${\bf r}_i$ than to any other point in the set. In two dimensions such a cell is a convex polygon, and the boundary of the Voronoi polygon is composed of segments of the perpendicular bisectors of each line (edge) that connects the point at ${\bf r}_i$ to its nearest-neighbor sites (points that share a Voronoi edge). The resulting Voronoi tessellation is then converted to a Voronoi network, by choosing the edges to be the intersections between two Voronoi cells and the nodes to be the locations where three cells intersect (i.e., we translate the points of the Voronoi tessellation into nodes and the lines into edges).  This allows us to easily compute the area $A_i$ of each Voronoi cell associated with the point at ${\bf r}_i$.


functions of the positions ${\bf r}_1$ and ${\bf r}_2$, e.g., knowing that there is a point in close proximity, very large areas are less
likely and the mean area decreases.

Following Ref. \cite{klatt2014characterization}, we define the Voronoi-area correlation function $C_{00}({\bf r}_1, {\bf r}_2)$ as the correlation between the volume of two Voronoi cells given that the corresponding centers are at the positions ${\bf r}_1$ and ${\bf r}_2$:
\begin{equation}
C_{00}({\bf r}_1, {\bf r}_2) = \displaystyle{\frac{<A({\bf r}_1)A({\bf r}_2)> - <A({\bf r}_1)><A({\bf r}_2)>}{\sigma({\bf r}_1 | {\bf r}_2)\sigma({\bf r}_2 | {\bf r}_1)}}
\label{eq_C00}
\end{equation}
where $<\cdot>$ denotes the ensemble average given two points at ${\bf r}_1$ and ${\bf r}_2$; and $\sigma({\bf r}_i | {\bf r}_j)$ is the standard deviation of the area of the Voronoi cell at ${\bf r}_i$ given that there is another point at ${\bf r}_j$. For a statistically homogeneous and isotropic system, the Voronoi-area correlation is simply a radial function of pair distance $r$, i.e., $C_{00}(r)$, where $r = |{\bf r}_2 - {\bf r}_1|$. The function $C_{00}(r) \in [-1, 1]$ measures the correlations, both positive and negative (i.e., anticorrelations), between Voronoi volumes of cells given that their centers are at a distance $r$.


We note that in general, $C_{00}(r)$ does not converge to perfect correlation for vanishing radial distance, i.e., $\lim_{r\rightarrow 0}C_{00}(r) <1$ because for all $r>1$, $C_{00}(r)$ provides the correlation of two different Voronoi cells with respectively areas $A(0)$ and $A(r)$. On the other hand, because the cell is perfectly correlated with itself, i.e., $C_{00}(r=0) = 1$, the correlation function $C_{00}(r)$ is discontinuous at the origin. If there is no long-range order, the correlation function tends to zero for infinite radial distance $\lim_{r\rightarrow \infty}C_{00}(r) = 0$.

\section{Results}


We generate the Voronoi tessellations of large, finite-sized two-dimensional systems containing $N\approx 10,000$ points in a large fundamental cell with periodic boundary conditions, which are built from hyperplane intersection process (HIP) \mbox{\cite{Chiu2013}}, Poisson, random sequential addition (RSA) \mbox{\cite{Torquato2006,Zhang2013}}, and both non-stealthy \mbox{\cite{wang2023equilibrium, wang2024designer}} and stealthy \mbox{\cite{Uc04,Ba08,To15}} point processes. Figure \ref{fig_1} shows small portions of representative Voronoi networks derived from these diverse set of point configurations. It can be clearly seen that sizes of Voronoi cells associated with the HIP and Poisson networks fluctuate significantly across the system; while the RSA and hyperuniform networks possess much more uniformly-sized Voronoi cells. We note that the area of Voronoi cells are related to the sizes of holes in the point pattern \cite{Zh15a}, which is subsequently connected to our current Voronoi network analyses.



\begin{table*}
\caption{Important statistics of Voronoi cell-area distribution of various systems, including the number of edges $M$ of the Voronoi tessellation, the first four cumulants (i.e., average $\bar{A}$, variance $\sigma^2_A$, skewness $\gamma_1$, and excess kurtosis $\gamma_2$), the information about the domain and range of the cell areas (i.e., the minimum area $A_{min}$, the maximum area $A_{max}$, the support $S = A_{max}-A_{min}$, the maximum probability density $P_{max}$ and the detail decay rate $\lambda_R$). }
\begin{ruledtabular}
\begin{tabular}{lccccccc}
&HIP & Poisson & RSA & \shortstack{Non-stealthy\\Hyperuniform} & \shortstack{Stealthy\\$\chi=0.20$} & \shortstack{Stealthy\\$\chi=0.40$} & \shortstack{Stealthy\\$\chi=0.49$}\\
\hline
$M$ & 100025 & 60004 & 100132 & 90000 & 100000 & 100000 & 100000\\
$\bar{A}$ & 1.00 & 1.00 & 1.00 & 1.00 & 1.00 & 1.00 & 1.00 \\
$\sigma^2_A$ & 1.60 & 0.283 & 0.0204 & 0.0580 & 0.0261 & 0.000633 & 0.000240\\
$\gamma_1$ & 3.16 & 1.07 & 0.450 & 0.257 & -0.327 & 0.0492 & 0.0272\\
$\gamma_2$ & 16.2 & 1.79 & -0.0383 & 0.0217 & 0.272 & -0.119 & -0.0626\\
$A_{min}$ &  4.88$\times10^{-6}$ & 0.0233 & 0.642 & 0.212 & 0.0770 & 0.896 & 0.940\\
$A_{max}$ & 19.5 & 5.64 & 1.73 & 2.22 & 1.57 & 1.11 & 1.06\\
S & 19.5 & 5.61 & 1.09 & 2.01 & 1.50 & 0.212 & 0.123\\
$P_{max}$ & 1.27 & 0.879 & 2.74 & 1.68 & 2.51 & 16.5 & 31.1\\
$\lambda_R$ & {0.237} & {1.00} & {9.93} & {3.93} & {13.2} & {57.8} & {127}
\end{tabular}
\end{ruledtabular}
\end{table*}

\subsection{Voronoi cell-area probability density distributions}

We carry out an extensive study of the Voronoi-cell area distribution of each of the networks. Figure \ref{fig_2} shows the normalized Voronoi cell-area probability density functions (PDF) calculated from Voronoi tessellations of approximately 10,000 faces associated with various systems. We normalize the cell areas by the density of the system $\rho$ such that the average area $\bar{A} = 1$ for all systems. A Gaussian PDF with the same mean and standard deviation of the corresponding Voronoi DPF is plotted on top of each distribution (red line). It can be clearly seen that the HIP networks possess PDFs that are distinctly different from the Gaussian distribution, while the other networks' PDFs appear to be well approximated by the Gaussian distribution.

To further quantify the cell-area distributions of the different networks, we compute multiple metrics that characterize the distribution, including first the four cumulants (i.e., average $\bar{A}$, variance $\sigma^2_A$, skewness $\gamma_1$, and excess kurtosis $\gamma_2$), the information about the domain and range of the cell areas (i.e., the minimum area $A_{min}$, the maximum area $A_{max}$, the support $S = A_{max}-A_{min}$, the maximum probability density $P_{max}$. We note that the excess kurtosis of the distribution is indicative of its tail behavior. Positive excess kurtosis indicates stronger tails (decaying slower) than a Gaussian distribution and negative excess kurtosis indicating weaker tails (decaying faster) than a Gaussian. The detailed values of these statistics are provided in Table 1.

Consistent with the visual inspection, We find that the antihyperuiform HIP distribution is far from Gaussian, as evidenced by a high skewness ($\gamma_1 = 3.16$) and large positive excess kurtosis ($\gamma_2 = 16.2$). The strong asymmetry of the HIP distribution is mainly due to the excess of large Voronoi cells, associated with the large holes in the system which also contribute to the large density fluctuations. The HIP distribution also possesses the largest variance $\sigma^2_A = 1.60$ and support $S = 19.5$ among all systems, indicating the extremely large variations and fluctuations of Voronoi cell sizes in the system.

The Poisson and RSA networks are both standard nonhyperuniform systems. As expected, the uncorrelated Poisson system possesses much larger variance $\sigma^2_A = 0.283$ and support $S = 5.61$ than the correlated RSA system with $\sigma^2_A = 0.0204$ and support $S = 1.09$. This indicates the Voronoi cells in the RSA networks are much more uniform in size, which is resulted from the uniform exclusion volume of the hard disks during the generation of the underlying point patterns. Moreover, the Poisson distribution is Gaussian-like, with a small and positive $\gamma_1 = 1.07$ and $\gamma_2 = 1.79$, except for its cutoff on the left because the area cannot be less than zero. On the hand, the RSA distribution is non-Gaussian, with $\gamma_1 = 0.450$ and $\gamma_2 = -0.0384$. The negative excess kurtosis indicates a fast-decaying detail than that in a Gaussian distribution. We note this is distinctly different non-Gaussian behavior than the HIP distribution. Narrower Voronoi cell-area distribution in the RSA system indicates that it is less likely for the system to contain very large the Voronoi cells (and thus, large holes).

Among the hyperuniform networks, the non-stealthy hyperuniform possesses the largest variance $\sigma^2_A = 0.0580$ and support $S = 2.01$. Both $\sigma^2_A$ and support $S$ decrease rapidly (and the peak value $P_{max}$ increases rapidly) as the degree of hyperuniformity (e.g., quantified via $\chi$) increases. The non-stealthy hyperuniform is also Gaussian-like, possessing small and positive skewness $\gamma_1 = 0.257$ and excess kurtosis $\gamma_2 = 0.0217$, which is consistent with previous result that this system can contain arbitrarily large holes in the thermodynamic limit \cite{wang2024designer}. On the other hand, the high-$\chi$ stealthy hyperuniform distributions (e.g., $\gamma_1 = 0.0272$ and $\gamma_2 = -0.0626$ for $\chi = 0.49$) are also non-Gaussian because of their low skewness and negative excess kurtosis, which is diametrically opposite non-Gaussian behavior of the HIP and indicates the fast decaying tail of the distribution.




\begin{figure*}[htpb]
    \includegraphics[width=0.75\textwidth]{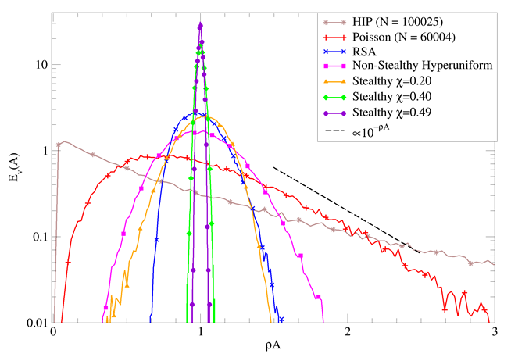}
\caption{Voronoi cell-area probability density functions for various system plotted on a semi-log axis. The scaling $ \propto 10^{-\rho A}$ is also shown as the dashed line. To allow for all the distributions to be discernible, we cutoff the x axes at 3. For visual clarity, we also only include symbols on every fifth data point.}
\label{fig_3}
\end{figure*}

\begin{figure*}[htpb]
    \includegraphics[width=0.95\textwidth]{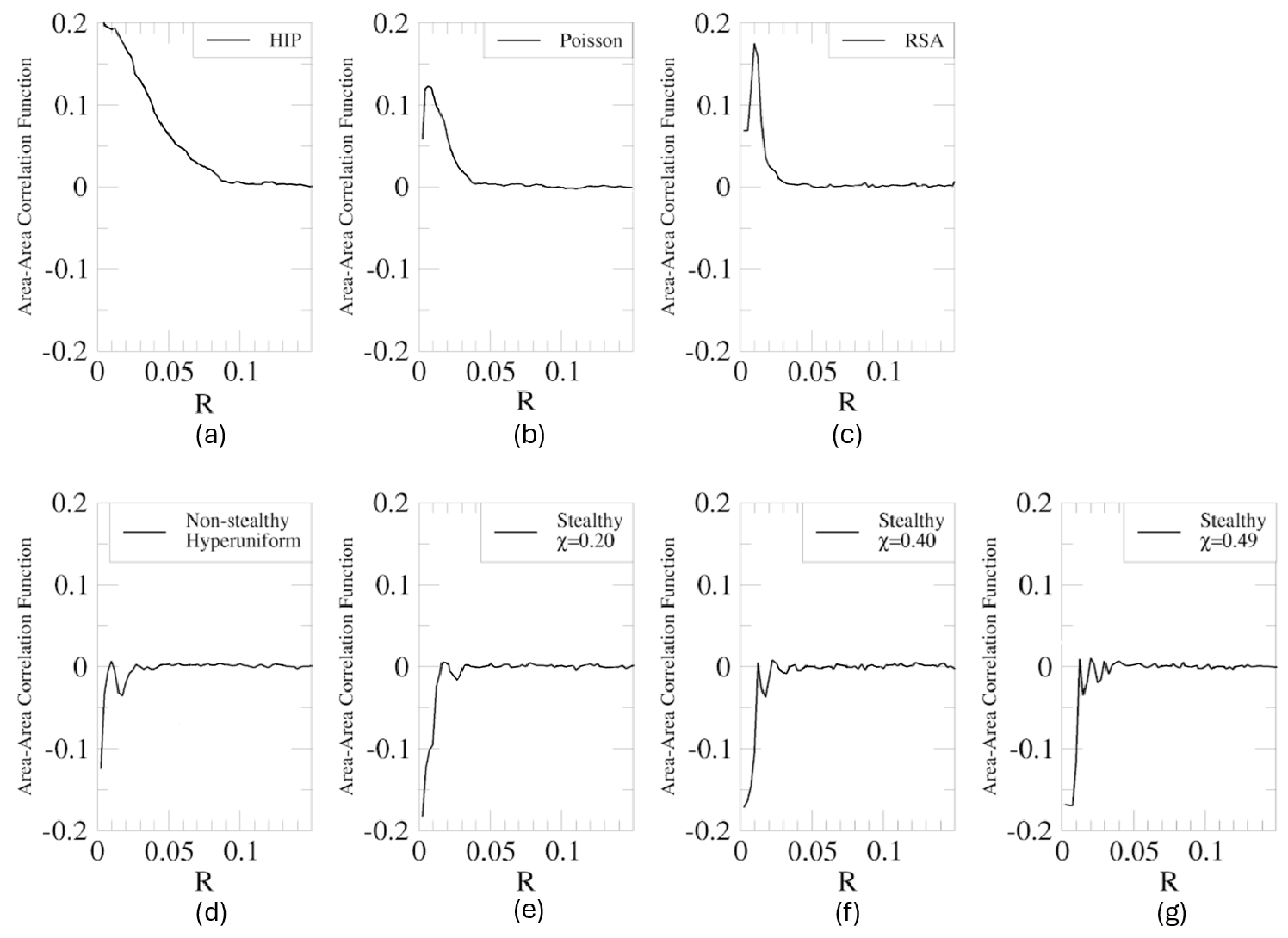}
\caption{Voronoi-area correlation functions $C_{00}(r)$ for the antihyperuniform HIP (a), nonhyperuniform Poisson (b) and RSA (c), non-stealthy hyperuniform (d), and stealthy hyperuniform networks with $\chi =0.2$ (e), $\chi =0.4$ (f) and $\chi =0.49$ (g). All hyperuniform networks exhibit strong anticorrelations in their Voronoi cell areas.}
\label{fig_4}
\end{figure*}

\subsection{Analysis of tail behavior of the distributions}

The fact that the cell-area distributions of these large, finite-sized RSA and stealthy hyperuniform networks (e.g., with $N \approx 10,000$ nodes) are narrower, have larger peaks and smaller tails than a Gaussian distribution implies that in the thermodynamic limit these distributions should exhibit compact support, consistent with previous theoretical considerations \cite{Zh17}. To further illustrate this point, we compare the tail behaviors of the distributions of all systems, which could be used to infer properties of support in the thermodynamic limit. We note that a system with infinite support in the thermodynamic limit possesses a distribution tail that decays at the rate of or slower than an exponential function. This behavior in principle can be captured in the analysis of large, finite-sized realizations as we have here.

In particular, Fig. \ref{fig_3} shows the Voronoi cell-area probability density functions for the different networks on a semi-log plot, which allows us to clearly distinguish the right tail behaviors of the different distributions. It can be clearly seen that the HIP, Poisson, and non-stealthy hyperuniform systems appear to possess a power-law tail, in contrast to the RSA and stealthy systems which appear to decreases (exponentially) faster.



To analyze the tail behavior of these systems, the slope of a linear regression of the end of the right tail is computed to approximate the decay rate of the right tail $\lambda_R$, see Table 1. To avoid the infinitely negative slope that every distribution must possess because of its finite nature, we measure the slope of the tail slightly before it intersects with the x-axis. While this calculation is only an approximation of decay rate of a non-exponentially decaying distribution, we can use $\lambda_R$ to observe how the networks decay relative to each other. This is reflected in the $\lambda_R$ values for our systems with the HIP, Poisson, and non-stealthy hyperuniform having small decay rates ($\lambda_R = 0.237, 1.00, 3.93$ respectively) while the RSA and stealthy systems have large decay rates ($\lambda_R = 9.93, 13.2, 57.8, 127$ for the RSA, stealthy $\chi = 0.20$, stealthy $\chi = 0.40$, and stealthy $\chi = 0.49$ respectively). From the large $\lambda_R$ values possessed by the RSA and stealthy systems we can infer that these systems have compact support in the infinite limit. On the other hand, the low $\lambda_R$ values of the HIP, Poisson, and non-stealthy hyperuniform allow us to infer infinite support for these systems in the thermodynamic limit. Overall, this analysis successfully recapitulates previously known results about these systems and expands on the current knowledge about the geometries of cells of the Voronoi tessellations of various point processes \cite{Andrage1988,Uc04,Kumar2005,Ferenc2007,Zh15a,Lazar2022}.




We note that the tail behavior of the Voronoi cell-area distributions is also connected to the hole statistics of the underlying point configurations. In particular, a \textit{hole} in the system of points constitutes a spherical region or radius $r$ that does not contain any points. The void exclusion probability function $E_V(r)$ (i.e., the probability that a hole of size $r$ exists in the infinite system) indicates if an arbitrarily large hole can exist in the system. For example, for an ideal gas of density $\rho$, the exclusion probability function is a decreasing exponential of the form:
\begin{equation}
    E_V(r) = e^{-\rho v_1(r)}
\end{equation}
Thus, arbitrarily large holes can exist for the ideal gas even if they have very small probability of occurring. Similarly, it has been found that non-stealthy hyperuniform point patterns allow for arbitrarily large holes \cite{Hough2009,Zh17,Ghosh2019}. Conversely, saturated random sequential addition (RSA) and stealthy hyperuniform point patterns have been proven to not tolerate arbitrarily large holes in the thermodynamic limit \cite{Zh17,Ghosh2018}.

\subsection{Voronoi-area correlation functions}


To further characterize the Voronoi networks of these distinct systems, we compute the Voronoi-area correlation functions $C_{00}(r)$ \cite{klatt2014characterization}. Figure \ref{fig_4} shows $C_{00}(r)$ for the antihyperuniform HIP, nonhyperuniform Poisson and RSA, non-stealthy hyperuniform, and stealthy hyperuniform networks. The antihyperuniform HIP networks possess a slowly decaying $C_{00}(r)$ with large positive values. We note that based on the definition (\ref{eq_C00}), larger deviation of $C_{00}(r)$ from 0 indicates larger fluctuations of the Voronoi cell areas. For a Voronoi network with identical cells
and orientations (i.e., one derived from a lattice), $C_{00}(r) = 0$ for all $r$ values. Therefore, the large positive values of $C_{00}(r)$ for antihyperuniform HIP networks indicates significant fluctuations of Voronoi cell areas across scales in the system. The nonhyperuniform Poisson and RSA networks also possess a positive $C_{00}(r)$ but with much faster decay.

On the other hand, all hyperuniform networks including both non-stealthy and stealthy hyperuniform ones exhibit strong anticorrelations in their Voronoi cell areas, manifested as the negative values of $C_{00}(r)$. Such anticorrelations of Voronoi cell areas indicate that in these networks large Voronoi cells are accompanied with small cells (and vice versa), which in turn suppresses large-scale area fluctuations to achieve hyperuniformity. The observed anticorrelations are consistent with previous analysis of the hyperuniform MRJ sphere packings in 3D \cite{klatt2014characterization}. Our results indicate that the correlation functions $C_{00}(r)$, which encode nonlocal pair-correlation information of Voronoi cells, can qualitatively distinguish the antihyperuniform, nonhyperuniform and hyperuniform Voronoi networks considered here.

\section{Conclusions and Discussion}


In this work, we present a comprehensive analysis of the structural properties of large two-dimensional nonhyperuniform and hyperuniform networks, derived from the Voronoi tessellations of antihyperuniform hyperplane intersection process, nonhyperuniform Poisson process, nonhyperuniform RSA saturated packings, and both non-stealthy and stealthy hyperuniform point processes. Specifically, We carried out an extensive study of the Voronoi-cell area distribution of each of the networks using multiple metrics that characterize the distribution, including their higher-cumulants (i.e., skewness $\gamma_1$ and excess kurtosis $\gamma_2$). We showed that the antihyperuniform HIP distribution is far from Gaussian, as evidenced by a high skewness ($\gamma_1 = 3.16$) and large positive excess kurtosis ($\gamma_2 = 16.2$). The nonhyperuniform Poisson (with $\gamma_1 = 1.07$ and $\gamma_2 = 1.79$) and non-stealthy hyperuniform (with $\gamma_1 = 0.257$ and $\gamma_2 = 0.0217$) distributions are Gaussian-like distributions as they exhibit a small but positive skewness and excess kurtosis. The nonhyperuniform RSA (with $\gamma_1 = 0.450$ and $\gamma_2 = -0.0384$) and the highest stealthy hyperuniform distributions (with $\gamma_1 = 0.0272$ and $\gamma_2 = -0.0626$) are also non-Gaussian because of their low skewness and negative excess kurtosis, which is diametrically opposite non-Gaussian behavior of the HIP. The fact that the cell-area distributions of large, finite-sized RSA and stealthy hyperuniform networks (e.g., with $N \approx 10,000$ nodes) are narrower, have larger peaks, and smaller tails than a Gaussian distribution implies that in the thermodynamic limit the distributions should exhibit compact support, which is consistent with previous theoretical considerations.


Moreover, we computed the Voronoi-area correlation functions $C_{00}(r)$ for the networks, which quantifies the correlations of areas of two Voronoi cells separated by distance $r$. We found that the antihyperuniform HIP networks possess a slowly decaying $C_{00}(r)$ with large positive values, while the nonhyperuniform Poisson and RSA network possess positive and fast decaying $C_{00}(r)$. In contrast, we observed strong anticorrelations in $C_{00}(r)$ (i.e., negative values) for the hyperuniform networks, including both the non-stealthy and stealthy ones. Such anticorrelations are realized by accompanying large Voronoi cells with small cells (and vice versa) in the systems, which suppressed large-scale area fluctuations to achieve hyperuniformity. These results indicate that $C_{00}(r)$ qualitatively distinguishes the antihyperuniform, nonhyperuniform and hyperuniform Voronoi networks considered here.


Our work here marks an attempt toward the generalization of analysis of hyperuniform point configurations, two-phase media and random fields to hyperuniform network systems. In general, a network can encode the complex interactions between individual elements of a system \cite{Bollobs1980,Newman2003,Watts2004,Barthelemy2011,barabasi2016,Newman2018} by building a topological map of the system that may or may not encode geometrical distance information. Networks have been used to model complex systems in many different fields, and their analysis has led to many fascinating discoveries \cite{Penrose2003,Barthelemy2011, jiao2012quantitative, Duchemin2023, Wasserman1994,Milo2002,Alm2003,Tyson2003,Alon2006,Vespignani2011,Albert2014,Papadopolous2018,Woerner2019,Aykol2019}. In our future work, we will further explore the existing network tools and develop new ones to investigate hyperuniform network systems.


\bigskip
\begin{acknowledgments}
This work was supported by the Army Research Office under Cooperative Agreement Number W911NF-22-2-0103.
\end{acknowledgments}


\end{document}